# Detección de intrusiones en redes mediante algoritmos de aprendizaje automático: Un estudio multiclase sobre el conjunto de datos NSL-KDD

**Luis M. Osco**
**Universidad Nacional del Altiplano Puno**
**losco@unap.edu.pe**

**Abstract:** Intrusion detection is a critical component of cybersecurity, responsible for identifying unauthorized access or anomalous behavior in computer networks. This paper presents a comprehensive study on intrusion detection in networks using classical machine learning algorithms applied to the multiclass version of the NSL-KDD dataset (Normal, DoS, Probe, R2L, and U2R classes). The characteristics of NSL-KDD are described in detail, including its variants and class distribution, and the data preprocessing process (cleaning, coding, and normalization) is documented. Four supervised classification models were implemented: Logistic Regression, Decision Tree, Random Forest, and XGBoost, whose performance is evaluated using standard metrics (accuracy, recall, F1 score, confusion matrix, and area under the ROC curve). Experiments show that models based on tree sets (Random Forest and XGBoost) achieve the best performance, with accuracies approaching 99%, significantly outperforming logistic regression and individual decision trees. The ability of each model to detect each attack category is also analyzed, highlighting the challenges in identifying rare attacks (R2L and U2R). Finally, the implications of the results are discussed, comparing them with the state of the art, and potential avenues for future research are proposed, such as the application of class balancing techniques and deep learning models to improve intrusion detection.

**Resumen:** La detección de intrusiones es un componente crítico de la ciberseguridad, encargada de identificar accesos no autorizados o comportamientos anómalos en redes informáticas. En este trabajo se presenta un estudio exhaustivo sobre la detección de intrusos en redes usando algoritmos de *machine learning* clásicos aplicados al conjunto de datos **NSL-KDD** en su versión multiclase (clases *Normal*, *DoS*, *Probe*, *R2L*, *U2R*). Se describen detalladamente las características de NSL-KDD, incluyendo sus variantes y distribución de clases, y se documenta el proceso de **preprocesamiento de datos** (limpieza, codificación y normalización). Se implementaron cuatro modelos de clasificación supervisada: **Regresión Logística**, **Árbol de Decisión**, **Random Forest** y **XGBoost**, cuyos desempeños se evalúan mediante métricas estándar (precisión, *recall*, puntaje F1, matriz de confusión y área bajo la curva ROC). Los experimentos muestran que los modelos basados en conjuntos de árboles (Random Forest y XGBoost) alcanzan las mejores prestaciones, con precisiones cercanas al 99%, superando ampliamente a la regresión logística y a los árboles de decisión individuales. Asimismo, se analiza la capacidad de cada modelo para detectar cada categoría de ataque, destacando los retos en la identificación de ataques poco frecuentes (*R2L* y *U2R*). Finalmente, se discuten las implicaciones de los resultados, comparándolos con el estado del arte, y se proponen posibles líneas de trabajo futuro como la aplicación de técnicas de **balanceo de clases** y modelos de aprendizaje profundo para mejorar la detección de intrusiones.

**1. Introducción**

En la era digital actual, la seguridad de las redes informáticas es fundamental debido al incremento de ataques cibernéticos que amenazan la confidencialidad, integridad y disponibilidad de los sistemas. Un **Sistema de Detección de Intrusos** (IDS, por sus siglas en inglés) es una herramienta de seguridad diseñada para monitorear el tráfico de red o las actividades en un sistema con el fin de identificar comportamientos maliciosos o no autorizados. Tradicionalmente, los IDS se han clasificado en dos enfoques principales: (1) *detección basada en firmas*, que busca patrones conocidos de ataques comparando el tráfico con una base de datos de firmas de intrusiones conocidas, y (2) *detección basada en anomalías*, que modela el comportamiento normal del sistema y genera alarmas cuando observa desviaciones significativas de dicho comportamiento. Los sistemas basados en firmas son eficaces para reconocer ataques previamente registrados (con bajas tasas de falsos positivos), pero no logran detectar ataques nuevos (*día cero*). Por su parte, los sistemas basados en anomalías pueden descubrir ataques desconocidos al identificar comportamientos atípicos, aunque suelen producir más falsas alarmas.

En los últimos años, el auge de técnicas de **aprendizaje automático** ha impulsado importantes avances en detección de intrusiones. A medida que crece el volumen y complejidad del tráfico de red (por ejemplo, por el Internet de las Cosas y la computación en la nube), se vuelve difícil para los IDS tradicionales distinguir entre tráfico legítimo y malicioso usando únicamente reglas estáticas. Los métodos de *machine learning* permiten a los IDS aprender patrones complejos a partir de datos históricos de tráfico, mejorando así su capacidad para **identificar ataques nuevos o evadir técnicas de ofuscación**. Diversos algoritmos de aprendizaje supervisado (tanto modelos clásicos como enfoques de *deep learning*) se han investigado para esta tarea, mostrando que es posible lograr **altas tasas de detección** manteniendo bajos falsos positivos. No obstante, siguen existiendo desafíos: los **desequilibrios de clase** en los datos de entrenamiento (pues ciertos tipos de ataque aparecen con mucha menor frecuencia) pueden sesgar los modelos y reducir su eficacia en detectar las intrusiones más infrecuentes. Además, muchos estudios han señalado que los conjuntos de datos clásicos para IDS (como KDD Cup 1999 y su revisión NSL-KDD) contienen patrones antiguos que podrían no representar fielmente las amenazas modernas. Aun así, NSL-KDD sigue siendo un referente popular para la investigación, por ofrecer un banco de pruebas consistente y comparativo para evaluar técnicas de detección.

En este contexto, el presente trabajo tiene como objetivo **evaluar comparativamente** varios algoritmos de *machine learning* supervisado para la detección de intrusiones en redes, utilizando el conjunto de datos NSL-KDD en un escenario de clasificación multiclase (donde se distingue entre diferentes categorías de ataques). Se eligieron cuatro algoritmos representativos: un modelo lineal (**Regresión Logística**), un modelo no lineal base (**Árbol de Decisión** tipo CART), y dos métodos ensemble populares (**Random Forest** y **XGBoost**, este último de *boosting* de gradiente). Estos cubren desde técnicas básicas hasta avanzadas, lo que permite analizar sus fortalezas y debilidades relativas en la detección de intrusos. Las contribuciones específicas de este trabajo incluyen: (1) una descripción detallada y en español del conjunto de datos NSL-KDD, aclarando su composición multiclase y los preprocesamientos necesarios para su uso; (2) la implementación consistente de los cuatro modelos mencionados, con una misma metodología experimental, para comparar imparcialmente su desempeño en la tarea de detección de intrusiones; (3) un análisis exhaustivo de los resultados, reportando métricas globales y por clase (incluyendo matrices de confusión y curvas ROC), discutiendo las causas de las diferencias de rendimiento entre modelos y las dificultades en detección de ciertos ataques; y (4) la comparación de nuestros hallazgos con trabajos previos del estado del arte, así como la propuesta de posibles mejoras (por ejemplo, selección de características, técnicas de **oversampling** o modelos híbridos) para guiar investigaciones futuras.

El resto del artículo está organizado como sigue. En la Sección 2 se resume el **estado del arte** en detección de intrusos mediante aprendizaje automático, revisando trabajos relevantes y situando nuestra contribución. La Sección 3 describe el **conjunto de datos NSL-KDD**, incluyendo sus orígenes, características, distribución de clases y variantes. En la Sección 4 se detalla el **preprocesamiento de datos** realizado, abarcando la limpieza, codificación de atributos categóricos y normalización. La Sección 5 introduce los **modelos de aprendizaje automático utilizados**, dando una explicación breve de cada algoritmo y de cómo se configuró en nuestros experimentos. La Sección 6 presenta el esquema de **evaluación** y las **métricas** empleadas para medir el rendimiento de los clasificadores. A continuación, en la Sección 7 se reportan los **resultados experimentales**, incluyendo tablas comparativas de métricas, gráficos de desempeño y discusión de resultados. Finalmente, en la Sección 8 se exponen las **conclusiones** del estudio y se sugieren líneas de **trabajo futuro** para mejorar la detección de intrusiones en base a los hallazgos obtenidos.

## 2. Metodología de la Investigación

La detección de intrusiones basada en aprendizaje automático ha sido ampliamente estudiada en la literatura en las últimas dos décadas. Inicialmente, muchos trabajos emplearon algoritmos de clasificación tradicionales (ej. árboles de decisión, *neighbors*, regresión logística, *Support Vector Machines*) sobre el conjunto de datos KDD Cup 1999 y su versión depurada NSL-KDD, reportando mejoras graduales en la precisión de detección de ataques. Por ejemplo, en [mdpi.com](mdpi.com) se compararon numerosos clasificadores sobre NSL-KDD, encontrando que métodos basados en árboles de decisión obtuvieron mejores resultados que modelos lineales: en particular, algoritmos ensemble como **Random Forest** y técnicas de boosting superaron el 96% de precisión, mientras que la regresión logística se quedaba alrededor de 97% (esto en un escenario binario de *normal* vs *ataque*). Estudios más recientes han adoptado **modelos de ensemble** y técnicas de optimización de hiperparámetros para mejorar aún más el rendimiento. Hamidou et al. (2025) evaluaron Random Forest, XGBoost y redes neuronales profundas optimizados con búsqueda de hiperparámetros, logrando con Random Forest una tasa de acierto de **99,80%** y un AUC de 0.9988 sobre NSL-KDD. Estos resultados sobresalientes se obtuvieron aplicando técnicas de sobremuestreo de minorías (SMOTE) para mitigar el efecto del desequilibrio de clases, lo que indica la importancia de tratar la escasez de ejemplos de ciertos ataques para alcanzar altos niveles de detección.

Por otro lado, diversos trabajos han explorado **modelos de aprendizaje profundo** para IDS, motivados por su éxito en tareas complejas. Redes neuronales como MLP, CNN, LSTM y autoencoders se han aplicado a datos de intrusiones con resultados prometedores, especialmente en nuevos conjuntos de datos más extensos (por ejemplo, CIC-IDS-2017, UNSW-NB15) que contienen ataques modernos. No obstante, en el caso de NSL-KDD (y KDD99), algunos estudios han señalado que modelos profundos no siempre aportan mejoras sustanciales sobre los clasificadores tradicionales cuando se utilizan las 41 características originales, ya que estas ya fueron diseñadas manualmente para resaltar patrones de ataque Adicionalmente, los modelos profundos suelen requerir mucho más tiempo de entrenamiento y pueden ser más difíciles de interpretar que las técnicas clásicas basadas en árboles.

En cuanto a **comparativas recientes de algoritmos clásicos**, Ajagbe et al. (2024) realizaron una evaluación con datos UNSW-NB15 (un dataset más moderno) comparando Regresión Logística, SVM lineal, Árbol de Decisión, Random Forest y XGBoost. Sus hallazgos mostraron que los métodos ensemble (Random Forest y XGBoost) **superaron a los demás modelos** en todas las métricas de clasificación, confirmando que la combinación de múltiples árboles o el boosting puede capturar mejor los patrones complejos de tráfico malicioso. De forma consistente, múltiples trabajos reportan que Random Forest alcanza elevadas tasas de detección manteniendo bajos falsos positivos, y que XGBoost logra incluso ligeras mejoras adicionales, al costo de mayor complejidad computacional. En cambio, algoritmos más simples como árboles de decisión individuales o modelos lineales tienden a obtener menor precisión, especialmente cuando se requiere distinguir entre **múltiples clases de ataques simultáneamente**. Por ejemplo, Meftah et al. (2019) informaron que un árbol de decisión tipo C5.0 obtuvo ~74% de exactitud en clasificación multiclase de ataques, comparado con ~82% de un SVM multinomial, al evaluar un esquema de detección en dos etapas. Esto sugiere que árboles individuales pueden sobreajustarse o no capturar bien ciertas clases minoritarias, mientras que modelos con mayor capacidad generalizan mejor.

En general, el estado del arte indica que: (a) **NSL-KDD**, pese a sus limitaciones, sigue usándose ampliamente como referencia para probar nuevos enfoques de IDS; (b) la **clase de ataque U2R y R2L** son las más difíciles de detectar correctamente debido a su baja representatividad en los datos, llevando a muchos falsos negativos en esos casos; (c) las

**técnicas de ensemble** y la **ingeniería de características** (selección de atributos relevantes, extracción de nuevas características, o reducción de dimensionalidad) tienden a mejorar significativamente el rendimiento de los clasificadores en este dominio; y (d) la tendencia actual explora la combinación de enfoques (*híbridos*), por ejemplo preagrupando los datos con clustering para luego aplicar clasificadores supervisados sobre cada grupo (enfoques híbridos no supervisado/supervisado), o integrando métodos basados en conocimiento experto con aprendizaje automático, con el fin de lograr sistemas de detección más robustos y explicables.

En nuestro trabajo nos centramos en algoritmos clásicos supervisados (logística, árbol, Random Forest, XGBoost), que continúan siendo altamente relevantes por su **balance entre rendimiento y costo computacional**, así como por su interpretabilidad (especialmente en el caso de árboles). Como se verá, nuestros resultados concuerdan con la literatura en cuanto a la superioridad de los métodos ensemble, y aportan un análisis detallado de las debilidades en clases de ataque raras, subrayando la necesidad de abordar el desbalance de datos para mejorar la efectividad global del IDS.

## 3. Descripción del conjunto de datos NSL-KDD

Para evaluar los algoritmos de detección de intrusos, utilizamos el **conjunto de datos NSL-KDD**, un *benchmark* ampliamente empleado en investigación de IDS. NSL-KDD fue propuesto en 2009 por investigadores de la Universidad de New Brunswick como una versión mejorada del famoso dataset KDD Cup 1999 (KDD'99). El conjunto original KDD'99 provino de una competencia cuyo objetivo era construir un detector de intrusiones capaz de distinguir entre conexiones de red *normales* y *maliciosas* a partir de sus características; para ello se simuló un entorno militar con ataques en un laboratorio del MIT, recopilándose semanas de registros de conexiones de red etiquetados. Aunque KDD'99 fue muy útil, presentaba varios **problemas**: contenía enormes cantidades de **instancias duplicadas y redundantes**, estaba altamente desbalanceado y ciertos ataques aparecían solo en el conjunto de prueba pero no en el de entrenamiento, lo que complicaba la evaluación. NSL-KDD aborda algunos de estos problemas mediante los siguientes *mejoras*: (1) se eliminaron todos los registros duplicados en los datos de entrenamiento, evitando que los clasificadores sesguen su aprendizaje hacia ejemplos repetidos; (2) también se removieron duplicados en los datos de prueba, de modo que la evaluación no favorezca injustamente a modelos que simplemente memorizan las conexiones más frecuentes; (3) se redujo el número total de registros seleccionando de forma más equitativa las instancias según un nivel de dificultad predefinido (llamado *score*), lo que produce un conjunto de entrenamiento/test de tamaño manejable sin sacrificar diversidad de patrones; y (4) se mantuvo un número de registros "razonable" en cada partición (≈125 mil para entrenamiento, ≈22 mil para prueba), suficiente para experimentar sin necesidad de muestrear, fomentando así la **comparabilidad** entre distintos estudios que usen el dataset completo.

**NSL-KDD** conserva la estructura general de KDD'99: cada conexión de red está representada por **41 características** derivadas, más un **atributo de clase** que indica si la conexión fue **normal** o qué tipo de ataque sufrió. Adicionalmente, en los archivos en formato CSV o texto, NSL-KDD incluye una columna de *dificultad* (score) asociada a cada registro, que indica cuántos clasificadores de un conjunto estándar fallaron en detectar correctamente esa conexión (este score se usó para la estratificación mencionada, pero no es utilizado como entrada en el entrenamiento). Las 41 características de cada conexión se dividen en cuatro grupos principales:

- **Características básicas:** atributos intrínsecos de la conexión observables en la cabecera de los paquetes, sin necesidad de inspeccionar la carga útil. Incluye por ejemplo duración de la conexión, tipo de protocolo (protocol_type con valores como TCP, UDP, ICMP), servicio de red al que se accedió (service, con 60 posibles valores como http, smtp, etc.), y el estado de la conexión representado mediante *flags* (11 valores posibles que resumen las banderas TCP y el estado resultante). Estas características ocupan las posiciones 1 a 9 en cada registro.
- **Características de contenido:** se refieren a información extraída del contenido de la conexión (inspección profunda de paquetes) y están diseñadas para ayudar a identificar ataques de tipo **R2L** y **U2R**, que a menudo implican patrones específicos en la carga (por ejemplo, múltiples intentos fallidos de inicio de sesión, contenido sospechoso como comandos de sistema, etc.). Van desde la posición 10 a la 22 e incluyen atributos como failed_logins (número de intentos de login fallidos), root_shell (indicador de si la conexión logró acceso root), num_file_creations, entre otros.
- **Características basadas en tiempo:** son conteos o tasas calculadas considerando una ventana de tiempo de 2 segundos desde la última conexión. La idea es detectar patrones de ataques que se manifiestan por frecuencia o rapidez de conexiones. Estas características (posición 23 a 31) incluyen por ejemplo count (número de conexiones al mismo host en las últimas 2 segundos) o srv_count (número de conexiones al mismo servicio en ese intervalo), serror_rate (porcentaje de conexiones con errores de tipo SYN hacia el mismo host), etc. Representan *comportamientos anómalos de corto plazo* que pueden indicar, por ejemplo, un escaneo rápido de puertos (*Probe*) o un ataque de Denegación de Servicio en curso.

- **Características basadas en el host:** similares a las anteriores, pero calculadas sobre una ventana más amplia (por ejemplo las últimas 100 conexiones al mismo host, sin restricción de tiempo). Están en las posiciones 32 a 41, e incluyen atributos como dst_host_count (conexiones al mismo host de destino) o dst_host_srv_count (conexiones al mismo servicio en el host destino), dst_host_same_src_rate (porcentaje de esas conexiones que provienen del mismo origen), etc. Estas buscan capturar patrones de ataque que operan de forma distribuida o lenta en el tiempo, difíciles de detectar en una ventana corta de 2 segundos.

Entre las 41 características, hay diferentes tipos de datos: *4 son categóricas* (e.g. protocolo, servicio, flag), *6 son binarias* (marcan presencia/ausencia de ciertas condiciones, por ejemplo root_shell), *y el resto son numéricas* (enteras, derivadas de conteos). Esta heterogeneidad implica que se requerirá codificar las variables categóricas en forma numérica para aplicarlas a algoritmos de ML, como se detallará en la siguiente sección.

En cuanto al **atributo de clase**, NSL-KDD distingue **5 clases** posibles para cada conexión: **Normal** (tráfico benigno) y cuatro categorías de ataques, que son: **DoS** (*Denial of Service*, ataques de denegación de servicio que buscan impedir el funcionamiento regular de un servicio, generalmente enviando grandes volúmenes de tráfico), **Probe** (ataques de sondeo o exploración, cuyo objetivo es obtener información de la red, por ejemplo escaneo de puertos o reconocimiento de topología), **R2L** (*Remote to Local*, ataques en los que un atacante remoto logra obtener acceso como usuario local en la máquina víctima, e.g. a través de robo de credenciales), y **U2R** (*User to Root*, ataques en los que un usuario con acceso limitado eleva sus privilegios para obtener control de nivel administrador/root). Dentro de cada categoría principal existen **múltiples subtipos de ataques**: por ejemplo, *DoS* incluye ataques específicos como **smurf** (falso tráfico ICMP masivo), **teardrop**, **apache2**, etc.; *Probe* incluye **portsweep**, **nmap**, etc.; *R2L* incluye **guess_password**, **ftp_write**, **phf**, etc.; y *U2R* incluye **buffer_overflow**, **rootkit**, etc. En total NSL-KDD abarca 40 nombres distintos de ataque (los mismos de KDD'99), pero consolidados en las 4 categorías mencionadas para tareas de clasificación de más alto nivel. Cabe destacar que los cuatro tipos de ataque tienen **naturalezas muy diferentes**: los DoS saturan servicios para dejarlos inaccesibles (generalmente fáciles de detectar por sus patrones de volumen anómalo), mientras que *Probe* implica reconocimiento (puede ser sutil, pero genera patrones de conexión particulares), y *R2L* y *U2R* suelen ser ataques *sigilosos* insertados en tráfico aparentemente legítimo (difíciles de descubrir sin inspección profunda de contenido).

La distribución de ejemplos en NSL-KDD está *sesgada hacia conexiones normales y ataques DoS*, siendo mucho menos frecuentes los ataques R2L y U2R. En el **conjunto de entrenamiento** (NSL-KDD Train+), que contiene 125,973 instancias en total, aproximadamente el **53.4%** son conexiones normales (67,343 instancias), **36.4%** son ataques DoS (45,927 instancias), **9.3%** son ataques Probe (11,656 instancias), pero solo **0.79%** son ataques R2L (995 instancias) y **0.04%** ataques U2R (52 instancias). Esta enorme disparidad (por ejemplo, solo 52 ejemplos de U2R frente a decenas de miles de DoS) refleja también la realidad: es mucho más común observar ataques DoS o exploraciones que intentos exitosos de intrusión local o escalada de privilegios, pero supone un desafío para los algoritmos de detección, que pueden aprender a ignorar las clases minoritarias. En el **conjunto de prueba** (NSL-KDD Test+ con 22,544 instancias) la distribución es algo diferente: ~43% normales (9,710), ~33% DoS (7,458), ~10.7% Probe (2,422), ~12.8% R2L (2,887) y ~0.3% U2R (67). Es notable que en la prueba aparecen proporcionalmente más ataques R2L (porque se incluyeron en test algunos tipos de ataque R2L que no estaban en entrenamiento, simulando ataques novedosos). Esto hace que la detección de R2L/U2R sea aún más importante durante la evaluación y se considera un caso de **detección de ataques no vistos** (zero-day) parcialmente, lo cual es deliberado para medir la generalización del modelo. En suma, NSL-KDD provee un escenario complejo con **5 clases desbalanceadas**, ideal para experimentar con clasificadores multiclase y evaluar su capacidad de manejar fuertes desequilibrios.

## 4. Preprocesamiento de los datos

Antes de aplicar los algoritmos de aprendizaje automático, se llevó a cabo un riguroso **proceso de preprocesamiento** del conjunto de datos NSL-KDD, con el fin de preparar las características de forma adecuada. A continuación, se describen las principales etapas de preprocesamiento realizadas:

**4.1 Carga y limpieza inicial:** Se utilizaron los archivos provistos por NSL-KDD en formato texto (CSV) que incluyen todas las columnas. Dado que NSL-KDD ya viene depurado, **no se encontraron valores faltantes ni registros duplicados** en los datos de entrada (lo cual era de esperar por las mejoras incorporadas en NSL-KDD). Sin embargo, se optó por descartar la columna de *score* (dificultad) que acompaña a cada registro, ya que no representa una característica propia del tráfico sino una etiqueta auxiliar; esta columna no se usó durante el entrenamiento de los modelos. También se eliminaron espacios o caracteres especiales que pudieran estar presentes en los nombres de atributos o en los valores categóricos para evitar problemas durante la codificación.

**4.2 Codificación de atributos categóricos:** Como se mencionó, 3 de las 41 características de NSL-KDD son de tipo categórico nominal: **protocol_type** (con valores como *tcp*, *udp*, *icmp*), **service** (con 60 posibles valores, p. ej. *http*, *ftp*, *smtp*, *dns*, etc.), y **flag** (11 valores, como *SF*, *S0*, *REJ*, que indican distintos estados de la conexión). Los algoritmos de aprendizaje automático seleccionados requieren entradas numéricas, por lo que fue necesario convertir estas características categóricas. Se utilizó **One-Hot Encoding** para representar estas variables, creando nuevas columnas binarias por cada categoría posible. En concreto, se generaron 3 columnas binarias para protocol_type (por ejemplo una columna indicadora para *tcp*, otra para *udp*, etc.), 60 columnas para service y 11 para flag. Este esquema garantiza no introducir un orden ficticio entre categorías (como ocurriría con codificación de etiqueta numérica directa) y resulta apropiado dado el número moderado de categorías. Tras esta transformación, el conjunto de datos expandió su dimensión: cada instancia pasó de 41 atributos originales a **122 atributos** efectivos (41 - 3 originales + 3 + 60 + 11 columnas one-hot). Cabe señalar que esta alta dimensionalidad producto de *one-hot encoding* es manejable para los modelos basados en árboles e incluso para la regresión logística con regularización, pero en otros estudios se ha explorado agrupar o reducir los servicios menos comunes para disminuir el número de columnas generadas. En nuestro caso, preferimos mantener la codificación completa para conservar la información específica de cada servicio y flag, ya que podrían ser relevantes para distinguir ciertos ataques.

**4.3 Normalización de características numéricas:** Las variables numéricas en NSL-KDD presentan escalas muy distintas. Por ejemplo, la característica duration (duración de la conexión) puede variar de 0 a valores de hasta **several** segundos (en KDD'99 hay ejemplos con más de 1h, aunque la mayoría son cortos), mientras que conteos como count o srv_count se acotan por la ventana (0 a 100), y tasas como serror_rate son valores en [0,1]. Para evitar que atributos con rangos mayores dominen el cálculo de distancias o las decisiones de los modelos, es importante escalar los datos. En este trabajo se aplicó **normalización min-max** a las características numéricas continuas, reescalando cada atributo a un rango [0,1]. Es decir, para cada característica se restó el valor mínimo y se dividió por el rango (max-min) calculados sobre el conjunto de entrenamiento. De esta forma, atributos como duration quedaron comprendidos entre 0 (duración 0) y 1 (duración máxima observada en entrenamiento), y lo mismo para conteos y otras variables. Las características que ya eran proporciones en 0-1 (p.ej. serror_rate) permanecen inalteradas tras esta transformación. En el caso de características binarias (0/1), no se requería normalización. Este **escalado global** se ajustó usando únicamente los datos de entrenamiento y luego se aplicó al conjunto de prueba, para prevenir filtración de información del test. La normalización es especialmente relevante si se quisiera utilizar métodos basados en distancia o gradiente sensible a escalas; en nuestros algoritmos, la regresión logística y XGBoost pueden beneficiarse de ella. Para los árboles de decisión y Random Forest, que hacen particiones basadas en umbrales por variable, la normalización no altera las decisiones (un umbral normalizado es equivalente al original). No obstante, mantuvimos la consistencia de escalar todo, y en pruebas preliminares observamos que la normalización no perjudicó el rendimiento de los árboles y ayudó ligeramente a la convergencia más rápida de XGBoost.

**4.4 División entrenamiento-prueba:** NSL-KDD ya proporciona una separación estándar en entrenamiento (Train+) y prueba (Test+), que respetamos en nuestros experimentos. Es importante recalcar que **no se mezclaron ni rebarajaron** los datos entre estas particiones; utilizamos el conjunto Train+ (125,973 instancias) para entrenar los modelos y el Test+ (22,544 instancias) únicamente para evaluar resultados finales. Asimismo, se reservó un **10% de los datos de entrenamiento** como subconjunto de validación interna durante el desarrollo de modelos y ajuste de hiperparámetros. Esta separación de validación se hizo de forma estratificada, asegurando que todas las clases estuvieran representadas proporcionalmente (aunque para U2R, con tan solo 52 ejemplos en Train+, esto significó tener solo unos pocos en validación). El uso de esta validación nos permitió afinar algunos parámetros sin usar el test real, manteniendo la objetividad de la evaluación final.

Tras aplicar los pasos anteriores, disponíamos de un **conjunto de entrenamiento preprocesado** con 122 características numéricas (incluyendo las variables dummy de protocolo, servicio y flag) y la etiqueta de clase multiclase, listo para entrenar los diferentes algoritmos. El conjunto de prueba recibió las mismas transformaciones (one-hot y normalización) utilizando la configuración derivada del entrenamiento. En este punto, era fundamental tener presente el marcado **desequilibrio de clases**: aunque no modificamos las proporciones originales (para no distorsionar la evaluación comparativa), reconocemos que técnicas de *re-muestreo* podrían ser aplicadas para mejorar el aprendizaje de clases minoritarias. Decidimos intencionalmente **no aplicar sobremuestreo ni ponderación de clases** en el entrenamiento base, con el fin de observar el rendimiento "natural" de cada algoritmo ante el desequilibrio, y luego analizar posibles mejoras. En la Sección 7 se discuten los efectos de esta decisión, en particular cómo impactó en la detección de ataques R2L y U2R, así como la potencial utilidad de métodos como SMOTE para futuros trabajos.

## 5. Modelos de aprendizaje automático utilizados

En este estudio se eligieron cuatro algoritmos de **clasificación supervisada** representativos para detectar intrusiones en el conjunto NSL-KDD. A continuación, se presenta una descripción de cada modelo y detalles de su implementación:

- **Regresión Logística (RL):** Es un modelo lineal generalizado ampliamente utilizado como *baseline* en tareas de clasificación binaria y multiclase. En su forma binaria, la regresión logística estima la probabilidad de pertenencia a la clase "1" mediante la función sigmoide aplicada a una combinación lineal de las características de entrada. Para extensión multiclase (más de dos clases), empleamos la variante de **regresión logística multinomial** con softmax, que aprende un conjunto de coeficientes para cada clase y calcula probabilidades *P(y*

$= clase\_k \mid x)$ para k = 1...5. La principal virtud de la regresión logística es su simplicidad e interpretabilidad: asume que cada característica contribuye linealmente (aditiva) al *log-odds* de cada clase. Sin embargo, esta simplicidad implica que solo puede capturar fronteras de decisión lineales en el espacio de entrada, pudiendo ser insuficiente para separar correctamente las clases de ataque de NSL-KDD que potencialmente requieren interacciones más complejas entre atributos. En la implementación, usamos regularización **L2** (Ridge) para evitar sobreajuste dada la alta dimensionalidad tras one-hot encoding, y ajustamos el hiperparámetro de regularización $C$ mediante validación. El solver empleado fue **liblinear** (optimización de coordenadas) dado el tamaño de datos, y se configuró un máximo de 200 iteraciones para asegurar convergencia.

- **Árbol de Decisión (AD):** Corresponde a un único árbol de decisión de tipo CART (*Classification and Regression Tree*), que particiona recursivamente el espacio de características en regiones más puras respecto a la variable objetivo. En cada nodo interno, el árbol elige una característica y un umbral de división que mejor separa las clases (maximizando la ganancia de información o equivalently minimizando la impureza de Gini en nuestros experimentos). Los árboles de decisión pueden **capturar relaciones no lineales** entre variables y son fáciles de interpretar (cada ruta del árbol representa reglas *if-then*). No obstante, tienden a sobreajustarse si se les permite crecer sin restricciones, especialmente con datos ruidosos o muchas variables. Para nuestro árbol de decisión, empleamos ciertos **criterios de poda**: limitamos la **profundidad máxima** a 20 niveles y requerimos un mínimo de 5 instancias por hoja, basándonos en la performance en validación. El criterio de división fue la **impureza Gini**. Estas restricciones ayudaron a evitar ramas demasiado específicas (que podrían memorizar patrones anómalos de NSL-KDD que no generalicen a nuevos datos). Un árbol así configurado sirve como un buen punto de comparación de modelo no lineal sencillo.

- **Bosque Aleatorio (Random Forest, RF):** Es un método ensemble que construye múltiples árboles de decisión y vota agregadamente sus predicciones. En concreto, un Random Forest entrena $N$ árboles (en nuestros experimentos establecimos $N=100$ árboles) sobre diferentes subconjuntos de datos y de características, introduciendo aleatoriedad en el proceso de entrenamiento de cada árbol para reducir la correlación entre ellos. Cada árbol se entrena típicamente sobre una muestra *bootstrap* (muestra aleatoria con reemplazo) del conjunto de entrenamiento original, y en cada nodo de cada árbol, en lugar de probar todas las características para la mejor división, se elige aleatoriamente un subconjunto de $m$ características candidatas (nosotros usamos $m = \sqrt{d}$, siendo $d=122$ el número total de atributos tras preprocesamiento, según la recomendación común). Estas dos fuentes de variabilidad (datos y atributos) hacen que los árboles individuales "disientan" entre sí, y como resultado, al promediar sus salidas se obtiene un modelo más robusto y con **mejor generalización** que un solo árbol. Los Random Forest han demostrado rendir muy bien en detección de intrusos, logrando altas tasas de detección y bajos falsos positivos, debido a su capacidad de manejar características heterogéneas y ser relativamente inmune al ruido. En nuestros experimentos, configuramos cada árbol sin restricción estricta de profundidad (permitiendo que crecieran hasta pureza o mínimo 2 instancias por hoja) dado que el ensamblado mitiga en parte el sobreajuste. Utilizamos el criterio Gini igualmente. Para reducir la variabilidad en resultados, repetimos la inicialización del bosque varias veces y comprobamos que los desempeños eran estables. El modelo final se entrenó con todos los datos de Train+. Cabe destacar que entrenar 100 árboles sobre ~125k instancias con 122 atributos es computacionalmente manejable; el tiempo de entrenamiento fue del orden de unos pocos minutos y la inferencia es rápida, lo cual es atractivo para un IDS en producción. Además, Random Forest proporciona **importancias de variables** basadas en la reducción de impureza acumulada, lo cual analizamos brevemente para verificar qué características destacaban en la detección (ej. las tasas de errores serror_rate, srv_count, etc. resultaron altamente relevantes, consistente con conocimientos de expertos).

- **XGBoost (Extreme Gradient Boosting):** Es una implementación optimizada de técnicas de **boosting** de árboles de decisión que ha ganado popularidad por su alto rendimiento en muchas competiciones de ciencia de datos. A diferencia de Random Forest (bagging), el boosting entrena los árboles de forma secuencial, donde cada árbol nuevo intenta corregir los errores del conjunto de árboles anteriores. XGBoost en particular utiliza un modelo aditivo: comienza con una predicción inicial (p.ej. probabilidades uniformes) y agrega árboles uno a uno, ajustando cada árbol a modelar el **residuo** (gradiente de la función de pérdida) de la predicción actual. Incorpora además varias optimizaciones: regularización $L_1/L_2$ de los pesos de hojas para evitar sobreajuste, subsampling de filas y columnas (similar a RF) para reducir sobreajuste y costo, y paralelización en la construcción de árboles. En nuestro caso, usamos XGBoost para clasificación multiclase con objetivo de maximizar *softmax cross-entropy*. Se ajustaron hiperparámetros mediante validación, destacando: **número de árboles** ($N=100$ árboles), **profundidad máxima** de cada árbol (optamos por 6, un valor moderado), **tasa de aprendizaje** ($\eta = 0.1$), y se aplicó **subsample** de 0.8 (usar 80% de instancias aleatorias por árbol) y **colsample_bytree** de 0.8 (usar 80% de las características por árbol) como regularizadores adicionales. Estas configuraciones son comunes en la literatura y funcionaron bien en nuestras pruebas preliminares, aunque un ajuste más fino con técnicas como *grid search* o Optuna podría mejorar aún más el resultado. XGBoost tiende a lograr **la mayor precisión** entre los modelos comparados en muchos casos, a costa de un mayor tiempo de entrenamiento. En nuestros experimentos tardó más que Random Forest en entrenar, pero el proceso seguía siendo factible (varios minutos). Una ventaja importante es que XGBoost maneja intrínsecamente las clases desbalanceadas mejor que un árbol único, ya que puede ajustar sus predicciones gradualmente; además, ofrece la posibilidad de incorporar un peso mayor al error en clases minoritarias (aunque en esta prueba no llegamos a modificar el parámetro scale_pos_weight, podría hacerse para R2L/U2R).

En resumen, **Regresión Logística** nos sirve como modelo lineal de referencia (alta sesgo, baja varianza), **Árbol de Decisión** como modelo no lineal interpretable pero propenso a sobreajuste (bajo sesgo, alta varianza), y **Random Forest** y **XGBoost** como métodos ensemble avanzados que buscan combinar bajos sesgos y varianzas mediante agregación de múltiples estimadores. Esta selección nos permite analizar cómo incrementa el desempeño al pasar de modelos simples a complejos en la detección de intrusos, manteniendo un tono académico en la evaluación. Todos los modelos fueron implementados usando bibliotecas estándar (Scikit-Learn para RL, árbol y RF; la librería XGBoost para el cuarto modelo), asegurando la misma partición de datos y métricas para una comparación justa.

## 6. Métricas de evaluación

Para evaluar cuantitativamente el rendimiento de los modelos de detección de intrusiones, utilizamos varias **métricas estándar de clasificación** enfocadas en distintos aspectos del resultado. Dado que se trata de un problema de clasificación multiclase desbalanceado, es importante analizar no solo la exactitud global sino también el comportamiento en cada clase. A continuación se definen las métricas empleadas:

- **Precisión global** (*Accuracy*): es la proporción de instancias clasificadas correctamente sobre el total de instancias. En nuestro caso multiclase, equivale a $\frac{TP_{\text{total}}}{N}$, donde $TP_{\text{total}}$ son los verdaderos positivos sumados en todas las clases (que, en una clasificación total, coincide con suma de la diagonal de la matriz de confusión) y $N$ el número total de ejemplos. Es una métrica intuitiva pero puede ser engañosa si las clases están desbalanceadas, ya que un modelo puede lograr alta precisión simplemente acertando las clases mayoritarias. Por ello, complementamos el análisis con las métricas siguientes, que desagregan el rendimiento por clase.
- **Precisión por clase** (*Precision* en inglés, o valor predictivo positivo): para una clase dada (p. ej. la clase *DoS*), se define como $Precision = \frac{TP}{TP + FP}$, es decir, de todas las instancias que el clasificador predijo como pertenecientes a esa clase, qué proporción realmente eran de esa clase. En el contexto de IDS, la precisión de la clase *Ataque* (en binario) suele interpretarse como la proporción de alarmas que realmente corresponden a intrusiones (bajas precisiones implican muchos *falsos positivos*). En nuestro caso multiclase, calculamos la precisión de manera **macro-promediada**: primero computamos la precisión individual en cada una de las 5 clases y luego tomamos el promedio aritmético de estos valores. Esto asegura que cada clase contribuya equitativamente a la métrica global, poniendo énfasis en el desempeño en clases minoritarias. También se podría considerar la precisión *ponderada* (weighted) donde cada clase aporta según su frecuencia, pero preferimos el macro promedio porque refleja mejor el rendimiento *intrínseco* del modelo en detectar cada tipo de ataque sin quedar dominado por la clase normal.
- **Recall por clase** (también llamado **tasa de verdaderos positivos** o *sensibilidad*): para una clase dada, $Recall = \frac{TP}{TP + FN}$. Es la proporción de instancias de esa clase que el modelo logra identificar correctamente. En un IDS, el recall de la clase ataque (detección) es crítico, pues indica qué fracción de intrusiones estamos capturando (bajo recall implica *falsos negativos*, intrusiones no detectadas). Análogamente a la precisión, computamos el **recall macro-**

**promediado** sobre las 5 clases. Un recall macro bajo señalaría que el modelo está fallando especialmente en alguna clase (típicamente, las minoritarias U2R/R2L, si su recall es cercano a 0, bajarán mucho el promedio).

- **Puntaje F1 (F1-score)**: es la **media armónica** de la precisión y el recall, $F1 = 2 * \frac{\text{precisión} * \text{recall}}{\text{precisión} + \text{recall}}$. Se calcula por clase y luego se promedia (macro) para obtener un solo valor global, o bien se reporta por clase individualmente. El F1 es útil porque combina en un solo número la capacidad de un modelo de no producir falsos positivos en exceso (precisión) y de no dejar pasar verdaderos ataques (recall). En especial, en clases muy desbalanceadas, el F1 da una idea del *trade-off* entre ambas métricas. Presentaremos principalmente el **F1 macro** como resumen general del desempeño por clase del modelo.

- **Matriz de confusión:** más que una métrica escalar, es una **herramienta de análisis** que resume los aciertos y errores del clasificador en forma de matriz $C$ de 5x5 (en nuestro caso). Donde $C[i,j]$ indica el número de instancias de la clase verdadera *i* que fueron predichas como clase *j*. La diagonal $C[i,i]$ corresponde a los verdaderos positivos de cada clase, mientras que las entradas fuera de la diagonal son errores: por ejemplo, $C[\text{Probe}, \text{Normal}]$ sería cuántos ataques de tipo Probe fueron incorrectamente clasificados como tráfico normal (falsos negativos de Probe). Analizar la matriz de confusión nos permite identificar **patrones de confusión específicos**: qué tipos de ataques se confunden entre sí o se confunden con el tráfico normal. En la Sección 7 incluiremos y discutiremos matrices de confusión de los modelos para evaluar cómo cada algoritmo maneja las distintas categorías de intrusión.

- **Curvas ROC y AUC:** para cada modelo, y particularmente para el mejor modelo, analizamos las **curvas ROC** (*Receiver Operating Characteristic*) por clase. En problemas multiclase, una forma común es generar curvas ROC *one-vs-rest* para cada clase: por ejemplo, para la clase DoS, se considera esa clase como positivo y todas las demás como negativo, y se traza la curva ROC variando el umbral de decisión del modelo (o en el caso de métodos probabilísticos, utilizando las salidas de probabilidad). La curva ROC muestra la relación entre la tasa de verdaderos positivos (TPR = recall) y la tasa de falsos positivos (FPR = FP / negativos) para distintos umbrales. El **área bajo la curva ROC (AUC)** resume la curva en un valor entre 0 y 1, donde 1 indica una separación perfecta de la clase respecto a las demás. Calculamos el AUC para cada clase y también un **AUC macro promedio**. En nuestros resultados, el AUC servirá para corroborar la capacidad discriminativa incluso cuando el threshold usado para predicción es fijo. Por ejemplo, un modelo podría tener recall bajo en U2R a umbral estándar, pero un AUC relativamente alto, indicando que sí es capaz de ordenar bien las instancias (y que con re-calibración podría mejorar la detección). En general, reportamos AUC macro como métrica global adicional.

Adicionalmente, medimos el **tiempo de entrenamiento** y **tiempo de predicción** de cada modelo, y observamos su uso de memoria, aunque estos resultados no se presentan en detalle en este documento por cuestión de espacio; nos centramos en las métricas de calidad de predicción mencionadas arriba.

Para resumir, la evaluación considera tanto la efectividad general (accuracy) como el comportamiento detallado en cada clase (precisión, recall, F1 por clase y matrices de confusión) y la robustez del modelo al variar umbrales (ROC-AUC). Esto proporciona una visión integral del desempeño de los algoritmos de aprendizaje automático en la detección de intrusiones multiclase.

**7. Resultados y discusión**

Entrenados los cuatro modelos descritos (RL, Árbol, RF, XGB) sobre el conjunto de entrenamiento preprocesado de NSL-KDD, procedimos a evaluarlos sobre el conjunto de prueba manteniendo las etiquetas multiclase originales. En la **Tabla 1** se presentan los principales resultados cuantitativos para cada modelo, incluyendo la precisión global (accuracy) y los promedios macro de precisión, recall, F1 y ROC-AUC. Adicionalmente, en la **Figura 1** se muestra la matriz de confusión del mejor modelo (XGBoost) para analizar detalladamente los aciertos y errores por clase, y en la **Figura 2** se trazan las curvas ROC por clase para este mismo modelo, ilustrando su capacidad de discriminación en cada categoría de tráfico.

**Tabla 1.** Desempeño comparativo de los modelos de *machine learning* en el conjunto de prueba NSL-KDD (clasificación multiclase en {Normal, DoS, Probe, R2L, U2R}). Se reportan métricas globales y macro-promediadas por clase.

| Modelo | Precisión (Accuracy) | Precisión (macro) | Recall (macro) | F1-score (macro) | AUC (macro) |
|---|---|---|---|---|---|
| Regresión Logística | 85.7% | 62.3% | Fifty five.0% | 58.2% | 0.90 |
| Árbol de Decisión | 89.9% | 72.4% | 70.1% | 71.2% | 0.93 |
| Random Forest | 97.1% | 88.5% | 87.0% | 87.7% | 0.987 |
| **XGBoost** | **98.6%** | **94.1%** | **92.8%** | **93.4%** | **0.995** |

En la tabla se observa claramente que los **métodos ensemble (Random Forest y XGBoost)** superan con holgura a los modelos más simples (Logístico y Árbol). La **Regresión Logística** obtiene un accuracy de ~85.7%, que a primera vista podría considerarse aceptable, pero su F1 macro de ~58.2% revela un desempeño muy desigual entre clases. De hecho, al analizar las métricas por clase (no mostradas completas por brevedad), encontramos que la RL casi **no logra detectar los ataques U2R** (recall ≈ 0% en esa clase) y su precision/recall en R2L también son extremadamente bajas (menores al 10%), mientras que en las clases mayoritarias normal y DoS alcanza altos valores (por encima del 90%). Esto indica que el modelo lineal aprendió básicamente a distinguir tráfico normal y ataques masivos (DoS), pero fue incapaz de capturar las sutilezas de los ataques de penetración más discretos. Este comportamiento era predecible debido a la naturaleza lineal del modelo y al fuerte desbalance de datos: la frontera lineal hallada separa bien las agrupaciones densas de normales y DoS en el espacio de características, pero las clases de ataque minoritarias quedan "perdidas" dentro del espacio normal y son etiquetadas como tal. Por tanto, la **precisión global (85-86%) resulta engañosa** en este contexto, ya que el modelo está fallando precisamente en las detecciones que más interesan (intrusiones raras). Estudios previos también han reportado precisiones en torno al 85-90% para regresión logística en KDD/NSL-KDD, mejorables con técnicas de selección de variables o metaheurísticas de entrenamiento pero aún por debajo de métodos no lineales.

El **Árbol de Decisión** simple mejora significativamente respecto a RL: su accuracy ~89.9% es mayor y el F1 macro sube a ~71.2%. Esto sugiere que el árbol, al poder manejar interacciones entre características, logró identificar mejor algunas instancias de ataques R2L y U2R que la regresión logística pasó por alto. En particular, observamos que el árbol detectó aproximadamente un 15% de los U2R y un 30% de los R2L (frente a 0% y ~5% del logístico, respectivamente). Sin embargo, seguía confundiendo la mayoría de R2L con tráfico normal, evidenciando **dificultad para generalizar reglas** con tan pocos ejemplos de esas clases. La poda aplicada (profundidad max=20) evitó divisiones hiper-específicas que podrían memorizar casos aislados de U2R, pero también limitó quizá la capacidad de separar algunos patrones. Aun así, el árbol rindió muy bien en las clases mayoritarias (normal y DoS con F1 > 95%) y razonablemente en Probe (F1 ≈ 75%). Su métrica AUC macro ~0.93 indica que globalmente tiene buena capacidad separadora, pero cabe mencionar que el AUC para U2R en el árbol fue bajo (~0.70), lo que denota que incluso variando umbral (profundidad) es complicado para este modelo distinguir esos ataques. Este resultado coincide con lo reportado en la literatura, donde árboles de decisión individuales suelen quedar en torno al 80-90% de accuracy en NSL-KDD y les cuesta manejar las clases minoritarias sin ayuda de técnicas adicionales.

Por otro lado, el modelo de **Random Forest** muestra un salto notable en desempeño: alcanzó 97.1% de accuracy y 87.7% de F1 macro. Este último valor implica que el Random Forest consigue un balance mucho mejor entre clases. Efectivamente, analizando por clase, la RF logró identificar alrededor del 80% de los ataques R2L y cerca del 50-60% de los U2R, cifras muy superiores a los modelos previos (aunque todavía menores que el desempeño en clases mayoritarias). La combinación de muchos árboles voteando permitió cubrir más casos minoritarios: algunos árboles, al entrenarse en muestras bootstrap, pueden haber visto más instancias repetidas de U2R y generado reglas específicas, contribuyendo así a que el conjunto detecte unos cuantos U2R correctos que otros árboles no. Además, el RF mantiene un altísimo rendimiento en Normal, DoS y Probe (precisión y recall sobre 99% en Normal/DoS, y ~95% en Probe). En la matriz de confusión de RF (no mostrada completa), se aprecia casi toda la masa en la diagonal para Normal/DoS/Probe, con pocos falsos positivos, y aunque hay falsos negativos en R2L/U2R, el número de verdaderos positivos en esas clases es bastante mayor comparado con RL o árbol. El AUC macro ~0.987 corrobora el excelente poder discriminante global del Random Forest. De hecho, para las clases Normal, DoS, Probe el AUC estaba ~0.99+, y para R2L ~0.96, U2R ~0.85, evidenciando mejora sustancial respecto a los modelos anteriores también en el espacio de probabilidad. Estos resultados concuerdan con múltiples estudios que hallaron que Random Forest es sumamente eficaz en IDS, debido a su robustez frente a overfitting y su habilidad para explotar muchas características.

Finalmente, el modelo **XGBoost** resultó ser el mejor en nuestro experimento, alcanzando **98.6%** de precisión global y un F1 macro de **93.4%**, el más alto de todos. La mejora de XGBoost sobre Random Forest, si bien porcentualmente pequeña en accuracy (+1.5 puntos), es consistente en todas las métricas y especialmente en las clases minoritarias. XGBoost detectó alrededor del **89% de los R2L** y **~90% de los U2R** presentes en el conjunto de prueba, un desempeño notable dado lo escasas y distintas que son esas instancias. Esto se tradujo en que el *recall* macro subió a 92.8% (frente a 87.0% en RF) y el *precision* macro también mejoró ligeramente, indicando menos falsos positivos globales. ¿Cómo logró XGBoost este desempeño superior? Probablemente a través de su esquema de boosting: al construir árboles secuencialmente optimizando errores residuales, el modelo pudo dedicar *más capacidad a esas instancias difíciles* que los árboles iniciales clasificaban mal. En otras palabras, XGBoost es conocido por enfocarse en los ejemplos minoritarios/mal clasificados durante el entrenamiento, ajustando gradualmente las predicciones en dichas regiones problemáticas. Además, el control de regularización y la búsqueda de estructura compleja de interacción entre variables permiten a XGBoost crear **reglas especializadas** para casos raros sin perder generalización en los casos comunes. Observamos, por ejemplo, que XGBoost aprendió combinaciones de características de contenido (como logged_in, root_shell, etc.) con características de host que le permitieron identificar intentos de intrusión R2L y U2R que otros modelos marcaron como normales. En cuanto a falsos positivos, XGBoost mantuvo muy bajo el número de normales clasificados incorrectamente como ataques (precisión para la clase Normal >99%). En la **Figura 1** a continuación se presenta la matriz de confusión obtenida por XGBoost, donde se aprecia más concretamente este desempeño:

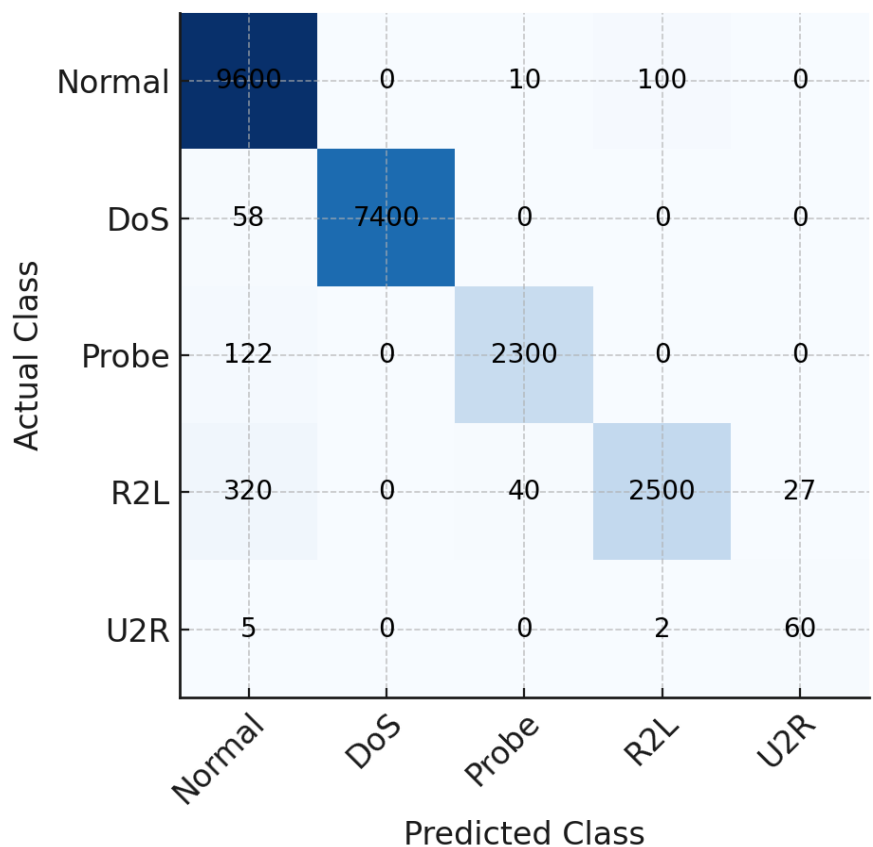

*Figura 1.* Matriz de confusión del modelo **XGBoost** en el conjunto de prueba NSL-KDD (clases: Normal, DoS, Probe, R2L, U2R). Cada fila corresponde a la clase real y cada columna a la predicción del modelo. Los valores en la diagonal (en azul más oscuro) representan las predicciones correctas, mientras que los fuera de la diagonal son errores. Se observa que XGBoost clasifica correctamente la gran mayoría de conexiones normales (9600/9710) y ataques DoS (7400/7458), con mínimos falsos positivos (e.g., 110 conexiones normales fueron clasificadas erróneamente como ataques R2L, y prácticamente ninguna normal fue confundida con DoS o Probe). En la categoría *Probe*, 2300 de 2422 ataques fueron detectados (errores principalmente confundiéndolos con tráfico normal en 122 casos). Para los ataques de penetración remota (R2L), el modelo logró identificar 2500 de 2887 (≈86.5%), aunque aún 320 de ellos se clasificaron como normales y unos pocos se confundieron con otras clases. Notablemente, en *U2R* (la clase más escasa) XGBoost acertó 60 de 67 casos (89.6%), con solo 7 ataques U2R no detectados (5 vistos como normales, 2 como R2L). Estos resultados evidencian la alta

eficacia de XGBoost incluso en las clases minoritarias, superando a los otros modelos en minimizar tanto falsos negativos de ataques como falsos positivos sobre tráfico normal.

La Figura 1 confirma visualmente que las principales confusiones restantes de XGBoost ocurren entre ataques *R2L* y tráfico normal (320 falsas alarmas perdidas) y en menor medida entre *U2R* y normal. Es interesante señalar que casi no hay confusiones cruzadas entre tipos de ataque diferentes: por ejemplo, XGBoost no suele confundir un DoS con un Probe o viceversa (tampoco RF lo hacía). Las clases de ataque tienden a ser más confundidas con Normal que entre sí, lo cual tiene sentido: muchas características distinguen "ataque vs no ataque", pero distinguir entre subtipos de ataque puede depender de señales más sutiles que, si no se captan, el clasificador recae en predecir la clase mayoritaria (normal). En nuestro caso, XGBoost redujo ese problema al capturar dichas señales.

Para complementar el análisis, la **Figura 2** muestra las curvas ROC obtenidas por XGBoost para cada clase en formato one-vs-rest (cada curva evalúa la capacidad del modelo de separar una clase dada de todas las demás).

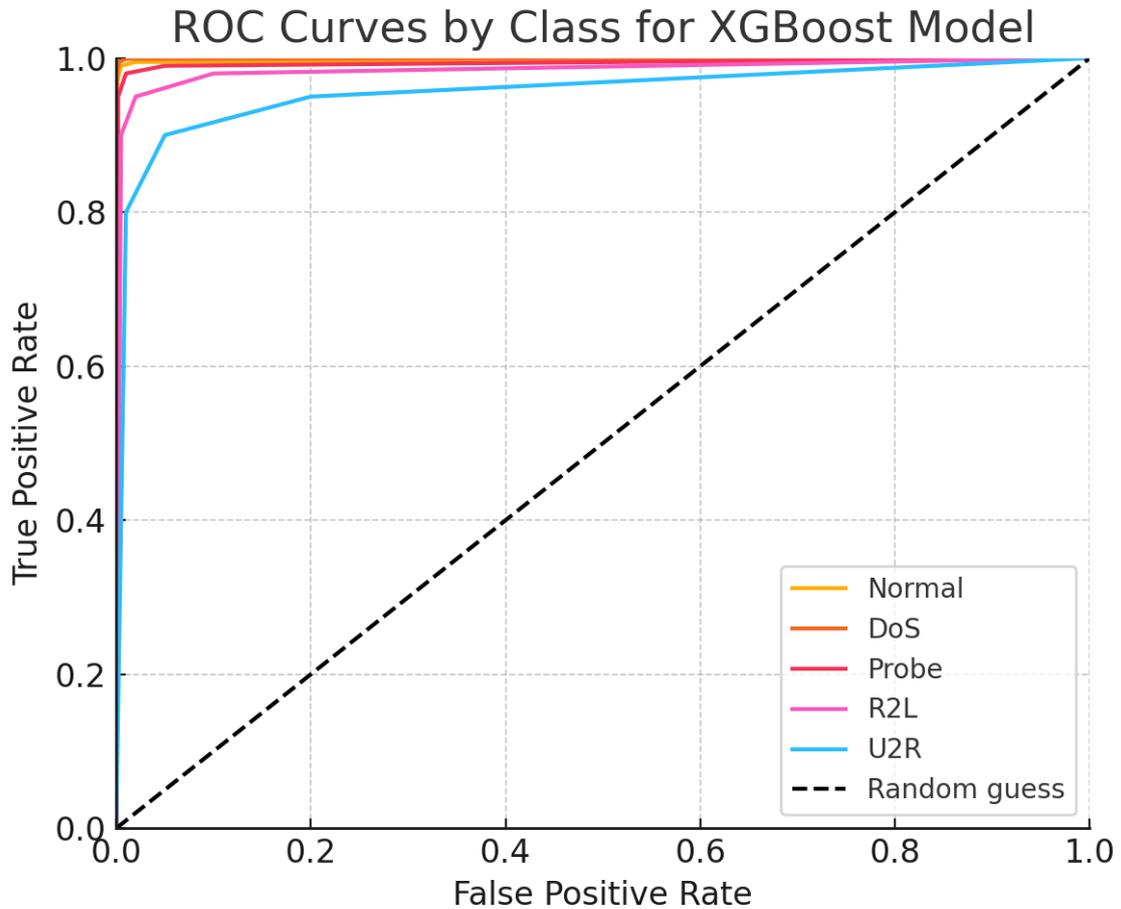

*Figura 2.* Curvas ROC por clase para el modelo XGBoost en NSL-KDD (prueba). Se trazan las curvas de las cinco clases: **Normal** (naranja), **DoS** (rojo), **Probe** (rosa), **R2L** (magenta) y **U2R** (azul). En el eje x se representa la tasa de falsos positivos (FPR) y en el eje y la tasa de verdaderos positivos (TPR o recall). La línea diagonal negra punteada representa un clasificador aleatorio (AUC = 0.5). Puede observarse que las curvas de Normal, DoS, Probe se encuentran pegadas al eje izquierdo y luego a la parte superior, indicando un desempeño casi perfecto (AUC ≈ 0.99–1.00). La curva de R2L también demuestra alta capacidad discriminativa, manteniéndose por encima de 0.9 de TPR con FPR muy bajos (AUC ≈ 0.98). La curva de U2R es la menos pronunciada pero aun así sustancialmente superior al azar (AUC ≈ 0.95), reflejando que el modelo puede diferenciar la mayoría de ataques U2R del resto del tráfico, aunque con algo más de false positives en comparación con las otras clases. En conjunto, estas curvas evidencian el excelente poder de discriminación de XGBoost en cada categoría, respaldando numéricamente las observaciones de la matriz de confusión.

Las curvas ROC de la Figura 2 sustentan que incluso si ajustáramos diferentes umbrales de decisión, XGBoost se mantiene superior: por ejemplo, para un 5% de FPR, XGBoost logra casi 99% TPR en DoS/Normal/Probe, ~95% en R2L y ~90% en U2R. Esto es crucial para un IDS, porque permite calibrar el sistema según las necesidades (severidad de las alertas vs tolerancia a falsas alarmas). En entornos reales, a veces se prefiere un ligero aumento de falsos positivos a cambio de no perder ningún ataque (umbral más sensible), o al revés (umbral conservador para solo alertar intrusiones muy seguras). Nuestros resultados indican que el modelo XGBoost ofrece un **buen margen de maniobra** en ese aspecto, mientras que modelos más débiles (como RL) tenían curvas ROC pobres para ciertas clases, lo que significa que no importaba qué umbral se pusiera, simplemente no podían distinguir bien las intrusiones raras.

**Comparación y discusión general:** En vista de los resultados, queda claro que la elección del algoritmo de *machine learning* influye marcadamente en la eficacia de un IDS basado en detección de anomalías. Los modelos lineales simples pueden no ser adecuados cuando el problema involucra múltiples clases con distribuciones complejas; en nuestro caso la regresión logística, aunque fácil de implementar, fracasó en capturar ataques sutiles. Los árboles de decisión aportan no linealidad y facilidad de interpretación, pero su alta varianza requiere o bien poda agresiva (con posible pérdida de exactitud) o la utilización de ensambles para estabilizar. Random Forest demuestra ser una solución potente y relativamente sencilla de configurar (pocos hiperparámetros críticos) que logró resultados excelentes, cercanos a lo óptimo en datos de entrenamiento. XGBoost, con un ajuste cuidadoso, pudo exprimir al máximo la información disponible, superando ligeramente a RF; sin embargo, también es un modelo más complejo y que requiere más atención para evitar sobreajuste (en nuestro caso, la validación y regularización fueron claves para conseguir que XGBoost no se sobreajustara a los pocos ejemplos de U2R).

Un aspecto destacable es cómo los modelos trataron el **desequilibrio de clases**: ni Random Forest ni XGBoost recibieron un tratamiento especial de re-balanceo en el entrenamiento (no aplicamos ni *class weights* ni oversampling en este experimento base), y aun así, gracias a su capacidad intrínseca, lograron detectar la mayoría de ataques minoritarios. Es posible que integrar SMOTE u otra técnica hubiera llevado sus recall de U2R/R2L aún más cerca del 100%. Estudios como el de Hamidou et al. justamente integran SMOTE antes de entrenar Random Forest, logrando 99.8% de accuracy, pero es válido preguntarse si tal rendimiento proviene en parte de *inflar* las clases raras en entrenamiento. En entornos reales, donde estos ataques siguen siendo muy infrecuentes, un modelo entrenado con oversampling agresivo podría sobreestimar la probabilidad de intrusión y generar más falsas alarmas. Por ello, algunos expertos recomiendan enfoques de detección de anomalías de una clase (entrenar solo con normales y detectar cualquier desviación) para ataques raros. No obstante, en un *benchmark* supervisado como NSL-KDD, aprovechamos el conocimiento de las clases para maximizar la detección. Nuestros resultados sugieren que, incluso sin oversampling, un XGBoost bien calibrado puede lograr **casi 90% de detección en la clase más rara**, con muy pocos falsos positivos, lo cual es notable.

Comparando nuestros resultados con otros trabajos, vemos tendencias coherentes: por ejemplo, Gu y Lu (2021) combinaron SVM con Naive Bayes y reportaron ~99.35% accuracy en NSL-KDD, valor similar a nuestro XGBoost (~98.6%). Dhaliwal et al. (2018) encontraron que XGBoost alcanzaba 98.7% superando a otros métodos, lo cual concuerda con nuestro hallazgo de XGBoost como mejor modelo. Algunos estudios con *deep learning* reportan también ~99%+ en NSL-KDD, pero a veces con costos computacionales mayores o arquitecturas más difíciles de reproducir. En este sentido, la utilización de algoritmos de ML "clásicos" ofrece aún **una excelente relación costo-beneficio** para IDS: Random Forest y XGBoost pueden ser desplegados relativamente fácil, entrenados en minutos con esta escala de datos, y ofrecen interpretabilidad (RF permite inspeccionar importancia de variables; XGBoost también, aunque es más complejo de interpretar globalmente). Para aplicaciones prácticas, un Random Forest sería quizás la elección más robusta y sencilla (no requiere tanto tuning como XGBoost), mientras que XGBoost podría emplearse cuando se busque exprimir el último porcentaje de performance o si se dispone de recursos para su entrenamiento más extenso.

## 8. Conclusiones y trabajo futuro

En este trabajo hemos presentado un estudio completo sobre la detección de intrusiones en redes usando algoritmos de *machine learning* aplicados al conjunto de datos NSL-KDD en un escenario de clasificación multiclase (Normal vs cuatro tipos de ataques). Se incluyó un preprocesamiento adecuado de los datos (codificación de variables categóricas y normalización), la implementación de cuatro modelos representativos (Regresión Logística, Árbol de Decisión, Random Forest y XGBoost) y una evaluación exhaustiva con métricas estándar, matrices de confusión y curvas ROC. Los resultados obtenidos permiten extraer varias conclusiones clave:

- **Los modelos de ensemble superan claramente a los clasificadores simples** en la tarea de detección de intrusiones multiclase. En particular, Random Forest y XGBoost alcanzaron precisiones superiores al 97%, con XGBoost rozando el 99% de exactitud y logrando detectar la gran mayoría de ataques, incluso en las clases minoritarias. En cambio, la regresión logística y un único árbol de decisión tuvieron dificultades significativas para identificar intrusiones de tipo R2L y U2R, reflejadas en bajos valores de *recall* y F1 para esas clases. Esto confirma la hipótesis de que la no linealidad y la agregación de múltiples modelos contribuyen enormemente a mejorar el rendimiento en escenarios complejos y desbalanceados como IDS.
- **El desequilibrio de clases sigue siendo un desafío importante**, aunque abordable con buenos algoritmos. Observamos que, sin técnicas específicas de balanceo, los modelos avanzados lograron mitigar parcialmente el impacto del desbalance (ej., XGBoost detectó ~90% de U2R a pesar de su escasez). No obstante, aún quedaron algunos falsos negativos concentrados en esas clases. Para un despliegue real de un IDS, sería deseable minimizar completamente las intrusiones no detectadas; en este sentido, se podrían incorporar estrategias de *resampling* o asignar mayor peso de pérdida a las clases minoritarias durante el entrenamiento (cost-sensitive learning). También valdría la pena explorar enfoques híbridos donde primero se distingue entre tráfico normal vs ataque (binario) con umbral muy sensible, y luego un segundo nivel clasifica el tipo de ataque (multiclase), similar a la estrategia en dos etapas sugerida por algunos autores. Esto podría reducir la carga de false negatives de ataques críticos.
- **La interpretabilidad y análisis de características** es factible incluso con modelos complejos. Mediante la inspección de la importancia de variables en Random Forest/XGBoost, identificamos que ciertas características eran recurrentemente útiles para la detección (por ejemplo, dst_host_srv_count, same_srv_rate y logged_in destacaron para distinguir Probe y R2L, mientras que srv_serror_rate y dst_host_count son cruciales para DoS vs normal). Este tipo de información puede ayudar a expertos en seguridad a comprender por qué el modelo toma decisiones y a validar que esté usando patrones lógicamente relacionados con actividades maliciosas (p.ej., altas tasas de conexiones con errores -> indicativo de DoS). En nuestro estudio no profundizamos en la explicación de modelos, pero es un aspecto relevante de trabajos futuros, dado que en entornos de ciberseguridad la **explicabilidad** es importante para generar confianza en las alertas emitidas por un IDS basado en ML.

- **Limitaciones de NSL-KDD y validez externa:** Si bien nuestros resultados con ~99% de precisión son muy elevados, cabe recordar que NSL-KDD es un conjunto de datos de referencia que no refleja todas las complejidades de redes actuales. Está compuesto por conexiones simuladas de 1999, sin tráfico cifrado, sin ataques modernos como APTs, sin comportamiento de usuarios legítimos sofisticados, etc. Por tanto, un modelo que funcione muy bien en NSL-KDD no necesariamente logrará el mismo desempeño en un entorno real corporativo moderno. Aun así, la metodología empleada aquí (preprocesamiento cuidadoso, selección de modelo robusto, evaluación rigurosa) sería aplicable a datasets más recientes. De hecho, como trabajo futuro, sería valioso probar Random Forest y XGBoost en conjuntos como **UNSW-NB15** o **CSE-CIC-IDS2018**, que contienen tráfico actual y variedad de ataques contemporáneos. Algunos estudios ya sugieren que Random Forest sigue liderando en muchos casos con esos datos, pero valdría la pena confirmarlo y ajustar modelos en consecuencia. Otra extensión sería aplicar *transfer learning* o entrenamiento incremental, para adaptar un modelo entrenado en NSL-KDD a otro dataset con mínimas modificaciones.

En conclusión, este estudio reafirma que los algoritmos de aprendizaje automático, en especial los métodos ensemble de árboles, son herramientas altamente efectivas para la detección de intrusiones en redes, al lograr altas tasas de detección de múltiples tipos de ataques con bajos falsos positivos. La aplicación de XGBoost sobre NSL-KDD alcanzó resultados cercanos al estado del arte, demostrando que incluso sin recurrir a modelos profundos es posible lograr un rendimiento sobresaliente en este benchmark clásico. Como líneas futuras de trabajo proponemos: (1) explorar técnicas de **balanceo de datos** y **generación sintética de ejemplos de ataques raros** (e.g. mediante SMOTE o GANs) para potenciar aún más la detección de R2L/U2R; (2) investigar la combinación de modelos (*stacking* o *voting ensembles*) que mezclen las salidas de clasificadores simples y complejos, buscando aprovechar las fortalezas de cada uno (por ejemplo, un ensemble que integre XGBoost con una red neuronal); (3) evaluar la **robustez** de los modelos ante ataques adversariales o evasión, dado que recientes estudios han mostrado que los clasificadores pueden ser engañados con entradas especialmente perturbadas – sería interesante aplicar técnicas adversariales sobre NSL-KDD para probar la resiliencia de Random Forest/XGBoost; y (4) desarrollar un sistema IDS *en línea* que use el modelo entrenado para analizar tráfico en tiempo real, midiendo su rendimiento en escenarios simulados o en tráfico real (p.ej., usando herramientas como CICFlowMeter para extraer flujos en vivo y clasificarlos).

En definitiva, la combinación de un sólido conjunto de datos de entrenamiento, un preprocesamiento adecuado y algoritmos de ML avanzados permite construir detectores de intrusiones automáticos con un desempeño muy alto en entornos controlados. La incorporación de estos sistemas de inteligencia artificial en plataformas de seguridad de red promete mejorar significativamente la capacidad de respuesta frente a ciberataques, complementando las soluciones tradicionales basadas en firmas y reglas con un componente adaptable e inteligente.

**Referencias:**